\documentclass[10pt,final,journal,twocolumn]{IEEEtran}
\ifCLASSINFOpdf

\else
\usepackage{algorithm}
\usepackage{booktabs}

\fi
\usepackage{cite}
\usepackage[cmex10]{amsmath}
\usepackage{amsthm}
\usepackage{stfloats}
\usepackage{multicol,multienum}
\usepackage{multirow}

\usepackage{amssymb}
\usepackage{url}
\usepackage{amsmath}
\usepackage{xcolor}
\usepackage[dvips]{graphicx}
\usepackage{subfigure}
\usepackage{caption}
\usepackage{amsmath}
\usepackage{amsfonts,amssymb}
\usepackage{bbm}
\usepackage[T1]{fontenc}
\usepackage{algorithm}
\usepackage{algpseudocode}
\usepackage{ulem}
\usepackage[linesnumbered, ruled]{}
\makeatletter
\renewcommand{\maketag@@@}[1]{\hbox{\m@th\normalsize\normalfont#1}}%
\makeatother

\begin{document}
% correct bad hyphenation here
\hyphenation{op-tical net-works semi-conduc-tor}
%\begin{document}
%\vspace{-2.75em}
\title{\LARGE \vspace{-0.8em} {Fluid Antennas}-Enabled Multiuser Uplink: A Low-Complexity Gradient Descent for Total Transmit Power Minimization}\vspace{-1em}
\author{Guojie Hu, Qingqing Wu,~\textit{Senior Member}, \textit{IEEE}, Kui Xu,~\textit{Member}, \textit{IEEE}, Jian Ouyang,~\textit{Member}, \textit{IEEE}, Jiangbo Si,~\textit{Senior Member}, \textit{IEEE}, Yunlong Cai,~\textit{Senior Member}, \textit{IEEE}, and Naofal Al-Dhahir,~\textit{Fellow}, \textit{IEEE}\vspace{-2.7em}
 %and Yunlong Cai,~\textit{Senior Member}, \textit{IEEE}
\thanks{
%Copyright (c) 2015 IEEE. Personal use of this material is permitted. However, permission to use this material for any other purposes must be obtained from the IEEE by sending a request to pubs-permissions@ieee.org.
%
%This work was supported in part by the Natural Science Foundations
%of China under Grants 62201606, 62071485, 62271503, 62071480 and 61971337, and in part by Natural Science Foundation of Jiangsu Province under Grant BK 20201334 and Shaanxi Province Natural Science Foundation
%for Distinguished Young Scholar (2022JC-50). (Corresponding author: Jiangbo Si)
%This work was supported in part by the Natural Science Foundations of China under Grants 62071480 and 61971337 and in part by the National Natural Science Foundation
%for Distinguished Young Scholar under Grant 61825104. (Corresponding author: Jiangbo Si)
This work was supported in part by the Natural Science Foundations
of China under Grants 62201606. Guojie Hu is with the College of Communication Engineering, Rocket Force University of Engineering, Xi'an 710025, China (email: lgdxhgj@sina.com). Qingqing Wu is with the Department
of Electronic Engineering, Shanghai Jiao Tong University, Shanghai 200240, China (qingqingwu@sjtu.edu.cn). Kui Xu is with the College of Communications Engineering, the Army of Engineering University, Nanjing 210007, China (lgdxxukui@sina.com). Jian Ouyang is with the Institute of Signal
Processing and Transmission, Nanjing University of Posts and Telecommunications, Nanjing 210003, China (ouyangjian@njupt.edu.cn). Jiangbo Si is with the Integrated Service Networks Lab of Xidian University, Xi'an 710100, China (jbsi@xidian.edu.cn). Yunlong Cai is with the College of Information Science and Electronic Engineering, Zhejiang University, Hangzhou 310027, China (email: ylcai@zju.edu.cn). Naofal Al-Dhahir is with the Department of Electrical and Computer Engineering, The University of Texas at Dallas, Richardson, TX 75080 USA (aldhahir@utdallas.edu).
%Guojie Hu and Donghui Xu are with the College of Communication Engineering, Rocket Force University of Engineering, Xi'an 710025, China (email: lgdxhgj@sina.com). Jiangbo Si and Zan Li are with the Integrated Service Networks Lab of Xidian University, Xi'an 710100, China (e-mail: jbsi@xidian.edu.cn, zanli@xidian.edu.cn). Yunlong Cai is with the College of Information Science and Electronic Engineering, Zhejiang University, Hangzhou 310027, China (email: ylcai@zju.edu.cn). Naofal Al-Dhahir is with the Department of Electrical and Computer Engineering, The University of Texas at Dallas, Richardson, TX 75080 USA (e-mail: aldhahir@utdallas.edu).
}%\vspace{-1.5em}
}
%This work was supported by the Natural Science Foundations of China (No. 61671474).
%L. X. Yang, D. Wu, and Y. M. Cai are with the College of Communications Engineering, the Army of Engineering University, Nanjing 210007, China. (Email: yanglianxin1228@126.com; wujing1958725@126.com; caiym@vip.sina.com.
\IEEEpeerreviewmaketitle
%\vspace{-5pt}
\maketitle
%\vspace{-20pt}
\begin{abstract}
We investigate multiuser uplink communications from multiple single-antenna users to a base station (BS), which is equipped with multiple fluid antennas (FAs) and adopts zero-forcing receivers to decode multiple signals. We aim to optimize antennas' positions at the BS, to minimize the total transmit power of all users subject to the minimum rate requirement. After applying transformations, we show that the problem is equivalent to minimizing the sum of each eigenvalue's reciprocal of a matrix, which is a function of all antennas' positions at the BS. Subsequently, the projected gradient descent (PGD) method is utilized to find a locally optimal solution. In particular, different from the latest related work, we exploit the eigenvalue decomposition to successfully derive a closed-form gradient for the PGD, which facilitates the practical implementation greatly. We demonstrate by simulations that via careful optimization for all antennas' positions in our proposed design, the total transmit power of all users can be decreased significantly as compared to competitive benchmarks.
\end{abstract}
\begin{IEEEkeywords}
Fluid antennas, multiuser uplink, total transmit power minimization, projected gradient descent.
\end{IEEEkeywords}

\IEEEpeerreviewmaketitle
\vspace{-5pt}
\section{Introduction}
Beamforming, which exploits the degree of freedom (DoF) in the spatial domain, is a powerful technique for improving system capacity [1]. In conventional beamforming, positions of antennas at transceivers are fixed which may limit the gains of beamforming depending on channel conditions.
%which leads to the static wireless channel environments. Such setting causes a fundamental limitation, i.e., in the case that the wireless channels are not ideal, beamforming may not be as useful as it should be.

To mitigate the above deficiency, the intelligent reflecting surface (IRS) technique has been proposed and proven to be capable of reconfiguring wireless channels by adjusting passive IRS reflecting coefficients [2]. As another promising technology, fluid antennas (FAs) [3]$-$[6] has emerged recently. Although its operating principle is different from that of the IRS, FAs can also reshape channel environments artificially, by adaptively adjusting positions of all antennas (connected to the radio frequency chains via flexible cables) supported by the stepper motors or servos. Unlike antenna selection (AS) which requires more candidate antennas, higher hardware cost and larger overhead of channel estimation, and concurrently unlike rotatable uniform linear array (RULA) which just mechanically rotates the transmit/receive array and cannot fully exploit spatial channel variation, FAs fully exploites the channel variation resulting from changes in antennas' positions to achieve a higher spatial diversity without causing additional hardware or algorithm cost [7].
%supported by the stepper motors or servos, the positions of all antennas (connected to the radio frequency chains via flexible cables) in the MA array can be adjusted in a specified region to artificially reshape the channel environments and then cater to effective beamforming as much as possible.
%
%
%, the MA technology does not increase the
 Driven by these potential advantages, earlier works have applied the technology of FAs to further enhance capacities of multiple-input multiple-output (MIMO) systems [7]$-$[8], multiuser uplink/downlink communications [9]$-$[10], physical-layer security systems [11] or interference networks [12].

 \begin{figure}
\centering
\includegraphics[width=5cm]{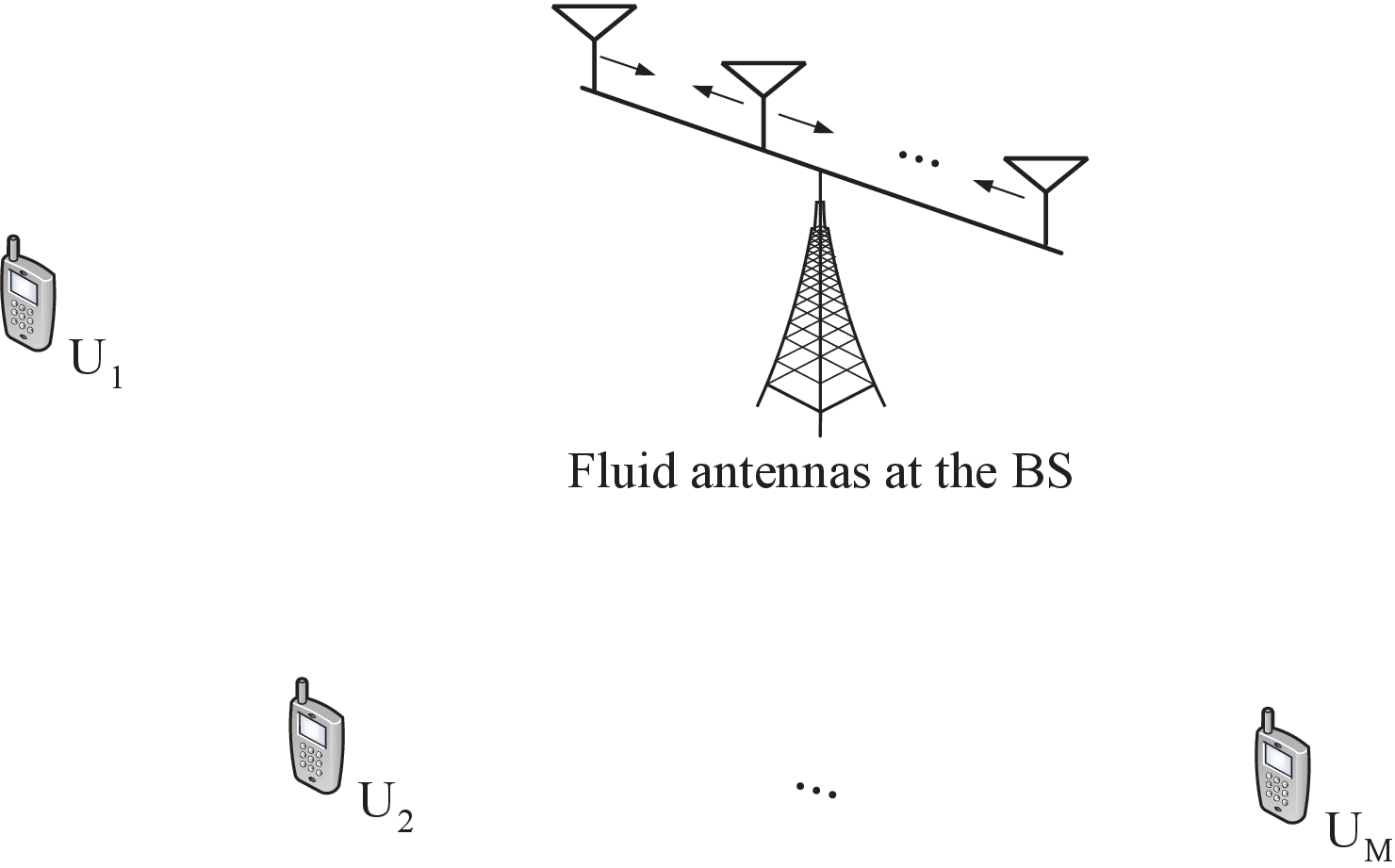}
\captionsetup{font=small}
\caption{Illustration of the system model.} \label{fig:Fig1}
\vspace{-15pt}
\end{figure}

In this letter, as in [9], we focus on FAs-enabled classical multiuser uplink communications. Specifically, we assume multiple single-antenna users that intend to concurrently transmit their signals to a base station (BS), which is equipped with FAs and adopts the widely used zero-forcing (ZF) receivers to detect multiple signals. By carefully optimizing positions of all antennas at the BS, our goal is to minimize the total transmit power of all users subject to the minimum rate requirement for each user. The formulated problem is highly non-convex, and we develop a projected gradient descent (PGD) method to find a locally optimal solution. Unlike [9] which exploits the original definition-based method to compute the gradient in each iteration, the key contribution of this letter is that we successfully derive a closed-form gradient in each iteration with the help of the eigenvalue decomposition. This novelty greatly accelerates the implementation of the PGD method. Numerical results are performed to demonstrate that our proposed method with FAs can significantly decrease the total transmit power of all users as compared to competitive benchmarks.

%\footnotemark \footnotetext{\textit{Notations:} ${{\bf{A}}^T}$, ${{\bf{A}}^H}$, ${{\bf{A}}^{ - 1}}$ and ${\rm{tr}}({\bf{A}})$ denote the transpose, conjugate transpose, inverse and trace of matrix ${\bf{A}}$, respectively. ${\rm{diag}}\left\{ {\bf{a}} \right\}$ is a diagonal matrix with the entry in the $i$-th row and $i$-th column equals the $i$-th entry of vector ${\bf{a}}$. ${\left[ {\bf{A}} \right]_{i,j}}$, ${\left[ {\bf{A}} \right]_{i,:}}$ and ${\left[ {\bf{A}} \right]_{:,i}}$ denote the entry in the $i$-th row and $j$-th column, $i$-th row vector and $i$-th column vector of matrix ${\bf{A}}$, respectively. $\left\| {\bf{a}} \right\|_2^2$ and $\left\| {\bf{A}} \right\|_{\rm{F}}^2$ denote the 2-norm of vector ${\bf{a}}$ and the Frobenius norm of matrix ${\bf{A}}$, respectively.}

 \newcounter{mytempeqncnt}
 \vspace{-5pt}
\section{System Model and Problem Formulation}
As shown in Fig. 1, we consider multiuser uplink communications from $M$ single-antenna users $\left\{ {{{\rm{U}}_i}} \right\}_{i = 1}^M$ to the BS equipped with $N$ FAs distributed along a linear dimension,
%\footnotemark \footnotetext{Actually, the BS can also employ the general planar array. However, we emphasize that the key conclusion of this letter is not affected by this extension. Thus, we consider a simpler setting for a clear presentation.}
with $N \ge M$. Consider the line-of-sight (LoS) propagation environment, the channel vector between the BS and ${{{\rm{U}}_i}}$ is denoted by\footnotemark \footnotetext{As shown in the follows, the simple LoS environment is considered here since we aim to demonstrate that our proposed design of FAs' movements just relies on the slow-changing property of statistical channel state information (CSI). In the simulations, we will show the effectiveness of the proposed FAs' movement rule when facing random Rician fading channels.}
{\setlength\abovedisplayskip{1.5pt}
\setlength\belowdisplayskip{1.5pt}
\begin{equation}
\begin{split}{}
{{\bf{h}}_i}({\bf{x}}) = {\left[ {{e^{j\frac{{2\pi }}{\lambda }{x_1}\sin {\theta _i}}},{e^{j\frac{{2\pi }}{\lambda }{x_2}\sin {\theta _i}}},...,{e^{j\frac{{2\pi }}{\lambda }{x_N}\sin {\theta _i}}}} \right]^T},
\end{split}
\end{equation}
where $\lambda $ is the signal wavelength, ${{\theta _i}}$ is the angle of arrival (AoA) to the BS at ${\rm{U}}_i$, and $x_n$ denotes the adjustable position of the $n$-th antenna at the BS, with ${\bf{x}} = {\left[ {{x_1},{x_2},...,{x_N}} \right]^T} \in {{\mathbb{R}}^{N \times 1}}$. For the multiuser uplink, the received signals ${\bf{y}} \in {{\mathbb{C}}^{M \times 1}}$ at the BS can be expressed as
\begin{equation}
\begin{split}{}
{\bf{y}} = {{\bf{W}}^H}{\bf{H}}({\bf{x}}){{\bf{P}}^{1/2}}{\bf{s}} + {{\bf{W}}^H}{\bf{n}},
\end{split}
\end{equation}
where ${\bf{H}}({\bf{x}}) = \left[ {{{\bf{h}}_1}({\bf{x}}),{{\bf{h}}_2}({\bf{x}}),...,{{\bf{h}}_M}({\bf{x}})} \right] \in {{\mathbb{C}}^{N \times M}}$, ${{\bf{P}}^{1/2}} = {\rm{diag}}\left\{ {\left[ {\sqrt {{P_1}} ,\sqrt {{P_2}} ,...,\sqrt {{P_M}} } \right]} \right\}$, in which $P_i$ denotes the transmit power of ${\rm{U}}_i$, ${\bf{s}} = {\left[ {{s_1},{s_2},...,{s_M}} \right]^T} \in {{\mathbb{C}}^{M \times 1}}$, in which $s_i$ denotes the transmitted signal of ${\rm{U}}_i$ and ${\mathbb{E}}\left[ {{{\left| {{s_i}} \right|}^2}} \right] = 1,\forall i = 1,...,M$. In addition, ${\bf{W}} = \left[ {{{\bf{w}}_1},{{\bf{w}}_2},...,{{\bf{w}}_M}} \right] \in {{\mathbb{C}}^{N \times M}}$ is the receive combining matrix at the BS, in which ${{\bf{w}}_i}$ is the combining vector for the signal $s_i$, and ${\bf{n}} = {\left[ {{n_1},{n_2},...,{n_N}} \right]^T}$, in which $n_i$ is the additive white Gaussian noise at the $i$-th BS antenna, with ${n_i} \sim {\cal C}{\cal N}(0,{\sigma ^2})$. Based on (2), the received signal-to-interference-plus-noise ratio (SINR) of the signal $s_i$ at the BS is derived as
\begin{equation}
\begin{split}{}
{\gamma _i} = \frac{{{P_i}{{\left| {{\bf{w}}_i^H{{\bf{h}}_i}({\bf{x}})} \right|}^2}}}{{\sum\nolimits_{k = 1,k \ne i}^M {{P_k}{{\left| {{\bf{w}}_i^H{{\bf{h}}_k}({\bf{x}})} \right|}^2} + \left\| {{{\bf{w}}_i}} \right\|_2^2{\sigma ^2}} }}.
\end{split}
\end{equation}

In this letter, we assume that the BS adopts the widely used linear ZF detector for processing multiple signals, due to its low implementation complexity especially when number of antennas at the BS is large. Based on this, the receive combining matrix ${\bf{W}}$ is accordingly expressed as
\begin{equation}
\begin{split}{}
{\bf{W}} = {\bf{H}}({\bf{x}}){\left( {{\bf{H}}{{({\bf{x}})}^H}{\bf{H}}({\bf{x}})} \right)^{ - 1}}.
\end{split}
\end{equation}
Substituting (4) into (3), the received SINR of the signal $s_i$ is given by
\begin{equation}
\begin{split}{}
{\gamma _i} = \frac{{{P_i}}}{{\left\| {{{\left[ {{\bf{H}}({\bf{x}}){{\left( {{\bf{H}}{{({\bf{x}})}^H}{\bf{H}}({\bf{x}})} \right)}^{ - 1}}} \right]}_{:,i}}} \right\|_2^2{\sigma ^2}}}.
\end{split}
\end{equation}

 Our goal is to optimize the positions of FAs at the BS, i.e., ${\bf{x}}$, to minimize the total transmit power of $M$ users subject to a minimum achievable rate requirement for each user. Hence, the optimization problem is formulated as\footnotemark \footnotetext{In this work, only antennas' positions are optimized for total power minimization. Consider the case where receiving beamforming and antennas' positions are jointly optimized, the generalized Bender's decomposition [13] can be exploited for obtaining the globally optimal solution.}
 \begin{align}
&({\rm{P1}}):{\rm{  }}\mathop {\min }\limits_{{\bf{x}},{\bf{P}}} \ \sum\nolimits_{i = 1}^M {{P_i}} \tag{${\rm{6a}}$}\\
{\rm{              }}&\ {\rm{s.t.}} \quad {\log _2}(1 + {\gamma _i}) \ge {r_i},\forall i = 1,...,M,\tag{${\rm{6b}}$}\\
&\quad \quad \ \ {\bf{x}} \in {\cal C},\tag{${\rm{6c}}$}
%& \ \ \ \quad {1_{\mathbb{C}}} = 1,\ {\rm{if}} \ \\
%&0 < {Q_E} \le \max \left( {{Q_{th}}/\left( {\frac{1}{{{\lambda _{EE}}}} + \frac{{{\beta %^r}N}}{{{\lambda _{RE}}{\lambda _{ER}}}}} \right),{Q_{E,\max }}} \right)\tag{${\rm{11d}}$}.
 \end{align}
where $r_i$ in the constraint (6b) denotes the minimum rate requirement for ${\rm{U}}_i$, and ${\cal C}$ in (6c) denotes the feasible moving region for $N$ antennas at the BS. More specifically, denote the total span for the movement of FAs as $L$ and without loss of generality set $0 \le {x_1} < {x_2} < ... < {x_N} \le L$. Then, consider: i) the minimum distance between any two FAs to avoid the coupling effect as ${{d_{\min }}}$ [7], [8], i.e., $\left| {{x_i} - {x_j}} \right| \ge {d_{\min }}$, $\forall i \ne j$; ii) the movement span should be the same for each antenna, we can conveniently set ${\cal C} \buildrel \Delta \over = \left\{ {{x_i} \in [{F_i},{G_i}]} \right\}_{i = 1}^N$, where
\begin{equation} \nonumber
\begin{split}{}
{F_i} =& \frac{{L - (N - 1){d_{\min }}}}{N}(i - 1) + (i - 1){d_{\min }},\\
{G_i} =& \frac{{L - (N - 1){d_{\min }}}}{N}i + (i - 1){d_{\min }},
\end{split}
\end{equation}
from which we have $0 = {F_1} < {G_1} < {F_2} < {G_2} < ... < {F_N} < {G_N} = L$ and ${G_i} - {F_i} = \frac{{L - (N - 1){d_{\min }}}}{N}$, $\forall i = 1,...,N$. The feasible movement region for each FA is illustrated in Fig. 2 for better understanding.

 \begin{figure}
\centering
\includegraphics[width=6cm]{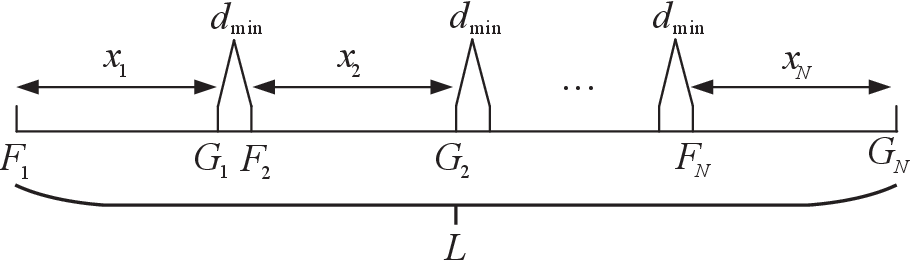}
\captionsetup{font=small}
\caption{Feasible movement region for each FA.} \label{fig:Fig1}
\vspace{-15pt}
\end{figure}

 \begin{figure*}[b!]
  \vspace{-0pt}
   \hrulefill
\setcounter{mytempeqncnt}{\value{equation}}
\setcounter{equation}{12}
\begin{equation}
\begin{split}{}
\frac{{\partial {\lambda _i}\left\{ {\bf{Z}} \right\}}}{{\partial {x_n}}} =& {\rm{Re}}\left[ {\frac{{\partial {{\left[ {{{\bf{V}}^{ - 1}}} \right]}_{i,:}}}}{{\partial {x_n}}}{\bf{Z}}{{\left[ {\bf{V}} \right]}_{:,i}} + {{\left[ {{{\bf{V}}^{ - 1}}} \right]}_{i,:}}\frac{{\partial {\bf{Z}}}}{{\partial {x_n}}}{{\left[ {\bf{V}} \right]}_{:,i}} + {{\left[ {{{\bf{V}}^{ - 1}}} \right]}_{i,:}}{\bf{Z}}\frac{{\partial {{\left[ {\bf{V}} \right]}_{:,i}}}}{{\partial {x_n}}}} \right]\\
\mathop  = \limits^{(a)} & {\rm{Re}}\left[ {\frac{{\partial {{\left[ {{{\bf{V}}^{ - 1}}} \right]}_{i,:}}}}{{\partial {x_n}}}{\lambda _i}\left\{ {\bf{Z}} \right\}{{\left[ {\bf{V}} \right]}_{:,i}} + {{\left[ {{{\bf{V}}^{ - 1}}} \right]}_{i,:}}\frac{{\partial {\bf{Z}}}}{{\partial {x_n}}}{{\left[ {\bf{V}} \right]}_{:,i}} + {\lambda _i}\left\{ {\bf{Z}} \right\}{{\left[ {{{\bf{V}}^{ - 1}}} \right]}_{i,:}}\frac{{\partial {{\left[ {\bf{V}} \right]}_{:,i}}}}{{\partial {x_n}}}} \right].
\end{split}
\end{equation}
\setcounter{equation}{\value{mytempeqncnt}}
\vspace{-5pt}
\end{figure*}

 \begin{figure*}[b!]
  \vspace{-0pt}
   \hrulefill
\setcounter{mytempeqncnt}{\value{equation}}
\setcounter{equation}{17}
\begin{equation}
\begin{split}{}
{\nabla _{{{\bf{x}}^t}}}f({\bf{x}}) = \left[ {\sum\nolimits_{i = 1}^M {\frac{{ - {{\left[ {{{\bf{V}}^{ - 1}}} \right]}_{i,:}}\frac{{\partial {\bf{Z}}}}{{\partial {x_1}}}{{\left[ {\bf{V}} \right]}_{:,i}}}}{{\lambda _i^2\left\{ {\bf{Z}} \right\}}}} ,\sum\nolimits_{i = 1}^M {\frac{{ - {{\left[ {{{\bf{V}}^{ - 1}}} \right]}_{i,:}}\frac{{\partial {\bf{Z}}}}{{\partial {x_2}}}{{\left[ {\bf{V}} \right]}_{:,i}}}}{{\lambda _i^2\left\{ {\bf{Z}} \right\}}}} ,...,\sum\nolimits_{i = 1}^M {\frac{{ - {{\left[ {{{\bf{V}}^{ - 1}}} \right]}_{i,:}}\frac{{\partial {\bf{Z}}}}{{\partial {x_N}}}{{\left[ {\bf{V}} \right]}_{:,i}}}}{{\lambda _i^2\left\{ {\bf{Z}} \right\}}}} } \right]_{{\bf{x}} = {{\bf{x}}^t}}^T.
\end{split}
\end{equation}
\setcounter{equation}{\value{mytempeqncnt}}
\vspace{-5pt}
\end{figure*}

Based on (6b), it can be shown that $P_i$ should satisfy
\begin{equation}
\setcounter{equation}{7}
\begin{split}{}
{P_i} \ge \left\| {{{\left[ {{\bf{H}}({\bf{x}}){{\left( {{\bf{H}}{{({\bf{x}})}^H}{\bf{H}}({\bf{x}})} \right)}^{ - 1}}} \right]}_{:,i}}} \right\|_2^2{\varepsilon _i}{\sigma ^2},
\end{split}
\end{equation}
where ${\varepsilon _i} = ({2^{{r_i}}} - 1)$. According to (7), we can equivalently replace the objective of (P1) as [9]
\begin{equation}
\begin{split}{}
&\sum\nolimits_{i = 1}^M {\left\| {{{\left[ {{\bf{H}}({\bf{x}}){{\left( {{\bf{H}}{{({\bf{x}})}^H}{\bf{H}}({\bf{x}})} \right)}^{ - 1}}} \right]}_{:,i}}} \right\|_2^2{\varepsilon _i}{\sigma ^2}} \\
 =& \left\| {{\bf{H}}({\bf{x}}){{\left( {{\bf{H}}{{({\bf{x}})}^H}{\bf{H}}({\bf{x}})} \right)}^{ - 1}}{{\bf{\Omega }}^{1/2}}} \right\|_{\rm{F}}^2\\
 =& {\rm{tr}}\left\{ {{{\left( {{{\bf{\Omega }}^{ - 1}}{\bf{H}}{{({\bf{x}})}^H}{\bf{H}}({\bf{x}})} \right)}^{ - 1}}} \right\}\\
 =& \sum\nolimits_{i = 1}^M {\frac{1}{{{\lambda _i}\left\{ {{{\bf{\Omega }}^{ - 1}}{\bf{H}}{{({\bf{x}})}^H}{\bf{H}}({\bf{x}})} \right\}}}} \buildrel \Delta \over = f({\bf{x}}),
\end{split}
\end{equation}
where ${\bf{\Omega }} = {\rm{diag}}\left\{ {\left[ {{\varepsilon _1}{\sigma ^2},{\varepsilon _2}{\sigma ^2},...,{\varepsilon _M}{\sigma ^2}} \right]} \right\}$ and ${{\lambda _i}\left\{ {{{\bf{\Omega }}^{ - 1}}{\bf{H}}{{({\bf{x}})}^H}{\bf{H}}({\bf{x}})} \right\}}$ denotes the $i$-th eigenvalue of the matrix ${{\bf{\Omega }}^{ - 1}}{\bf{H}}{({\bf{x}})^H}{\bf{H}}({\bf{x}}) \buildrel \Delta \over = {\bf{Z}} \in {{\mathbb{C}}^{M \times M}}$. Therefore, problem (P1) can be equivalently reformulated as
\begin{align}
&({\rm{P2}}):{\rm{  }}\mathop {\min }\limits_{{\bf{x}}} \ f({\bf{x}})  \tag{${\rm{9a}}$}\\
{\rm{              }}&\ {\rm{s.t.}} \quad {\bf{x}} \in {\cal C}.\tag{${\rm{9b}}$}
%& \ \ \ \quad {1_{\mathbb{C}}} = 1,\ {\rm{if}} \ \\
%&0 < {Q_E} \le \max \left( {{Q_{th}}/\left( {\frac{1}{{{\lambda _{EE}}}} + \frac{{{\beta %^r}N}}{{{\lambda _{RE}}{\lambda _{ER}}}}} \right),{Q_{E,\max }}} \right)\tag{${\rm{11d}}$}.
 \end{align}

\textbf{Remark 1:} Problem (P2) is highly non-convex because its objective is neither convex or concave, which cannot be solved via standard convex optimization techniques. Motivated by this, the authors in [9] try to solve (P2) by resorting to the PGD method, which handles the simple unconstrained or constrained problems well and is not sensitive to concavity or convexity of the objective. However, [9] computes the gradient based on the original definition shown in its equation (12), which has the large implementation complexity. In the next section, we show how to reduce the complexity significantly.

\textbf{Remark 2:} Considering the LoS environment, the BS can easily estimate the CSI by just estimating the AoAs to itself at $M$ users based on some mature algorithms, such as MUSIC. Based on this, the BS can directly optimize FAs' positions via the proposed algorithm and then feedback each user the required transmit power based on (7) with optimized ${\bf{x}}$.}\footnotemark \footnotetext{In addition, even the general Rician fading is considered, the BS still optimizes FAs' positions in advance based on the estimated AoAs. Then, in the communication process, all antennas' positions are not changed and each user sends pilot signals to the BS for uplink channel estimations. When the BS successfully estimates the instantaneous CSI, it can tell each user the required transmit power based on (7). Since no antennas' movements are involved, the consumed time for estimate-feedback is much smaller than the channel coherence time (CCT), especially for the low-mobility scenario where CCT is relatively larger [14].}

\section{Algorithm Design for Solving (P2)}
In this letter, we still exploit the PGD method to find a locally optimal solution to (P2). Specifically, using PGD, the update rule for ${\bf{x}}$ in the $t + 1$-th iteration is given by
{\setlength\abovedisplayskip{1.5pt}
\setlength\belowdisplayskip{1.5pt}
\begin{equation}
\setcounter{equation}{10}
\begin{split}{}
{{\bf{x}}^{t + 1}} =& {{\bf{x}}^t} - \delta {\nabla _{{{\bf{x}}^t}}}f({\bf{x}}),\\
{{\bf{x}}^{t + 1}} =& {\cal B}\left\{ {{{\bf{x}}^{t + 1}},{\cal C}} \right\},
\end{split}
\end{equation}
where ${{\bf{x}}^{t + 1}}$ in the first equation is the original updated ${\bf{x}}$, and ${{\bf{x}}^{t + 1}}$ in the second equation is the additional update (if necessary) via the projection function ${\cal B}\left\{  \cdot  \right\}$ as explained later, which ensures that the solutions for FAs' positions in each iteration always satisfy the constraint in (9b). Further, ${\nabla _{{{\bf{x}}^t}}}f({\bf{x}})$ denotes the gradient of $f({\bf{x}})$ at ${{{\bf{x}}^t}}$, and $\delta $ is the step size for the gradient descent.
\vspace{-0.1pt}
%\subsection{Computing ${\nabla _{{{\bf{x}}^t}}}f({\bf{x}})$}
\vspace{-0.1pt}

{\textit{\underline{A. Computing ${\nabla _{{{\bf{x}}^t}}}f({\bf{x}})$:}}} Note that ${\nabla _{\bf{x}}}f({\bf{x}}) = {\left[ {\frac{{\partial f({\bf{x}})}}{{\partial {x_1}}},...,\frac{{\partial f({\bf{x}})}}{{\partial {x_N}}}} \right]^T}$. Using the chain rule, $\frac{{\partial f({\bf{x}})}}{{\partial {x_n}}}$, $\forall n = 1,...,N$, can be derived as
\begin{equation}
\begin{split}{}
\frac{{\partial f({\bf{x}})}}{{\partial {x_n}}} = \sum\nolimits_{i = 1}^M {\frac{{ - 1}}{{\lambda _i^2\left\{ {\bf{Z}} \right\}}}} \frac{{\partial {\lambda _i}\left\{ {\bf{Z}} \right\}}}{{\partial {x_n}}}.
\end{split}
\end{equation}
Based on (11), to compute ${\nabla _{\bf{x}}}f({\bf{x}})$, the key is to derive a closed-form expression for $\frac{{\partial {\lambda _i}\left\{ {\bf{Z}} \right\}}}{{\partial {x_n}}}$, $\forall i = 1,...,M$ and $n = 1,...,N$.

To proceed, let us denote ${\bf{Z}} = {\bf{VD}}{{\bf{V}}^{ - 1}}$ as the eigenvalue decomposition of the matrix ${\bf{Z}}$, where ${\bf{V}} \in {{\mathbb{C}}^{M \times M}}$ consists of linearly independent columns with unit norm, and ${\bf{D}} = {\rm{diag}}\left\{ {\left[ {{\lambda _1}\left\{ {\bf{Z}} \right\},...,{\lambda _M}\left\{ {\bf{Z}} \right\}} \right]} \right\}$. Then, we can equivalently express ${\lambda _i}\left\{ {\bf{Z}} \right\}$ as
\begin{equation}
\setcounter{equation}{12}
\begin{split}{}
{\lambda _i}\left\{ {\bf{Z}} \right\} = {\left[ {{{\bf{V}}^{ - 1}}} \right]_{i,:}}{\bf{Z}}{\left[ {\bf{V}} \right]_{:,i}}.
\end{split}
\end{equation}
Based on (12), $\frac{{\partial {\lambda _i}\left\{ {\bf{Z}} \right\}}}{{\partial {x_n}}}$ can be expanded as in (13), where $\mathop  = \limits^{(a)} $ is established since ${\bf{Z}}{\left[ {\bf{V}} \right]_{:,i}} = {\lambda _i}\left\{ {\bf{Z}} \right\}{\left[ {\bf{V}} \right]_{:,i}}$ and ${\left[ {{{\bf{V}}^{ - 1}}} \right]_{i,:}}{\bf{Z}} = {\lambda _i}\left\{ {\bf{Z}} \right\}{\left[ {{{\bf{V}}^{ - 1}}} \right]_{i,:}}$. Then, further note that the sum of the first and third terms in (13) equals
\begin{equation}
\setcounter{equation}{14}
\begin{split}{}
&\frac{{\partial {{\left[ {{{\bf{V}}^{ - 1}}} \right]}_{i,:}}}}{{\partial {x_n}}}{\lambda _i}\left\{ {\bf{Z}} \right\}{\left[ {\bf{V}} \right]_{:,i}} + {\lambda _i}\left\{ {\bf{Z}} \right\}{\left[ {{{\bf{V}}^{ - 1}}} \right]_{i,:}}\frac{{\partial {{\left[ {\bf{V}} \right]}_{:,i}}}}{{\partial {x_n}}}\\
 =& {\lambda _i}\left\{ {\bf{Z}} \right\}\frac{{\partial \left[ {{{\left[ {{{\bf{V}}^{ - 1}}} \right]}_{i,:}}{{\left[ {\bf{V}} \right]}_{:,i}}} \right]}}{{\partial {x_n}}}\mathop  = \limits^{(b)} 0,
\end{split}
\end{equation}
where $\mathop  = \limits^{(b)} $ is established since ${{{\left[ {{{\bf{V}}^{ - 1}}} \right]}_{i,:}}{{\left[ {\bf{V}} \right]}_{:,i}}}$ always equals the constant one and thus is not relevant to $x_n$ in any situation. Based on (13) and (14), $\frac{{\partial {\lambda _i}\left\{ {\bf{Z}} \right\}}}{{\partial {x_n}}}$ can be simplified as
\begin{equation}
\begin{split}{}
\frac{{\partial {\lambda _i}\left\{ {\bf{Z}} \right\}}}{{\partial {x_n}}} =&{\rm{Re}}\left[ {{{\left[ {{{\bf{V}}^{ - 1}}} \right]}_{i,:}}\frac{{\partial {\bf{Z}}}}{{\partial {x_n}}}{{\left[ {\bf{V}} \right]}_{:,i}}} \right]\\
 \mathop  = \limits^{(c)} & {{{\left[ {{{\bf{V}}^{ - 1}}} \right]}_{i,:}}\frac{{\partial {\bf{Z}}}}{{\partial {x_n}}}{{\left[ {\bf{V}} \right]}_{:,i}}},
\end{split}
\end{equation}
%
%\begin{equation}
%\begin{split}{}
%&\frac{{\partial {\lambda _i}\left\{ {\bf{Z}} \right\}}}{{\partial {x_n}}} = {\rm{Re}}\left[ {{{\bf{V}}^{ - 1}}(i,:)\frac{{\partial {\bf{Z}}}}{{\partial {x_n}}}{\bf{V}}(:,i)} \right]\\
% =& {\rm{Re}}\left[ {{{\bf{V}}^{ - 1}}(i,:)\frac{{\partial \left[ {{{\bf{\Omega }}^{ - 1}}{\bf{H}}{{({\bf{x}})}^H}{\bf{H}}({\bf{x}})} \right]}}{{\partial {x_n}}}{\bf{V}}(:,i)} \right]\\
% =& {\rm{Re}}\left[ {{{\bf{V}}^{ - 1}}(i,:){{\bf{\Omega }}^{ - 1}}\frac{{\partial \left[ {{\bf{H}}{{({\bf{x}})}^H}{\bf{H}}({\bf{x}})} \right]}}{{\partial {x_n}}}{\bf{V}}(:,i)} \right]
%\end{split}
%\end{equation}
%
%Recall that ${{\bf{V}}^{ - 1}}(i,:){\bf{ZV}}(:,i) = {{\bf{V}}^{ - 1}}(i,:){{\bf{\Omega }}^{ - 1}}{\bf{H}}{({\bf{x}})^H}{\bf{H}}({\bf{x}}){\bf{V}}(:,i) = {\lambda _i}({\bf{Z}})$. Further, same as ${\bf{H}}{({\bf{x}})^H}{\bf{H}}({\bf{x}})$, ${\frac{{\partial \left[ {{\bf{H}}{{({\bf{x}})}^H}{\bf{H}}({\bf{x}})} \right]}}{{\partial {x_n}}}}$ is still a hermite matrix. Hence, ${{{\bf{V}}^{ - 1}}(i,:){{\bf{\Omega }}^{ - 1}}\frac{{\partial \left[ {{\bf{H}}{{({\bf{x}})}^H}{\bf{H}}({\bf{x}})} \right]}}{{\partial {x_n}}}{\bf{V}}(:,i)}$ is a real number, which finally results in
%\begin{equation}
%\begin{split}{}
%\frac{{\partial {\lambda _i}\left\{ {\bf{Z}} \right\}}}{{\partial {x_n}}} = {{\bf{V}}^{ - 1}}(i,:)\frac{{\partial {\bf{Z}}}}{{\partial {x_n}}}{\bf{V}}(:,i)
%\end{split}
%\end{equation}
where $\mathop  = \limits^{(c)} $ is established since ${{{\left[ {{{\bf{V}}^{ - 1}}} \right]}_{i,:}}\frac{{\partial {\bf{Z}}}}{{\partial {x_n}}}{{\left[ {\bf{V}} \right]}_{:,i}}}$ is a real number. Recall that ${\bf{Z}} = {{\bf{\Omega }}^{ - 1}}{\bf{H}}{({\bf{x}})^H}{\bf{H}}({\bf{x}})$ and ${{\bf{\Omega }}^{ - 1}} = {\rm{diag}}\left\{ {\left[ {1/({\varepsilon _1}{\sigma ^2}),1/({\varepsilon _2}{\sigma ^2}),...,1/({\varepsilon _M}{\sigma ^2})} \right]} \right\}$. The element in the $i$-th row and $j$-th column of ${\bf{Z}}$ based on (1) can be derived as
\begin{equation}
\begin{split}{}
{\left[ {\bf{Z}} \right]_{i,j}} = \frac{1}{{{\varepsilon _i}{\sigma ^2}}}\sum\nolimits_{k = 1}^N {{e^{j\frac{{2\pi }}{\lambda }{x_k}\left( {\sin {\theta _j} - \sin {\theta _i}} \right)}}},
\end{split}
\end{equation}
based on which it is easy to derive the element in the $i$-th row and $j$-th column of $\frac{{\partial {\bf{Z}}}}{{\partial {x_n}}}$ as
\begin{equation}
\begin{split}{}
&{\left[ {\frac{{\partial {\bf{Z}}}}{{\partial {x_n}}}} \right]_{i,j}} = \frac{{\partial {{\left[ {\bf{Z}} \right]}_{i,j}}}}{{\partial {x_n}}} \\
=& \frac{1}{{{\varepsilon _i}{\sigma ^2}}}\frac{{2\pi }}{\lambda }\left( {\sin {\theta _j} - \sin {\theta _i}} \right){e^{j\left[ {\frac{{2\pi }}{\lambda }{x_n}\left( {\sin {\theta _j} - \sin {\theta _i}} \right) + \frac{\pi }{2}} \right]}}.
\end{split}
\end{equation}
Finally, by substituting the known ${\frac{{\partial {\bf{Z}}}}{{\partial {x_n}}}}$ into (15) and then substituting (15) into (11), the gradient ${\nabla _{\bf{x}}}f({\bf{x}})$ at ${{\bf{x}}^t}$ can be computed as in (18).
%\begin{equation}
%\begin{split}{}
%&{\nabla _{{{\bf{x}}^t}}}f({\bf{x}}) = \left[ \begin{array}{l}
%\sum\nolimits_{i = 1}^M {\frac{{ - {{\bf{V}}^{ - 1}}(i,:)\frac{{\partial {\bf{Z}}}}{{\partial {x_1}}}{\bf{V}}(:,i)}}{{\lambda _i^2\left\{ {\bf{Z}} \right\}}}} ,...,\\
%...,\sum\nolimits_{i = 1}^M {\frac{{ - {{\bf{V}}^{ - 1}}(i,:)\frac{{\partial {\bf{Z}}}}{{\partial {x_N}}}{\bf{V}}(:,i)}}{{\lambda _i^2\left\{ {\bf{Z}} \right\}}}}
%\end{array} \right]_{{\bf{x}} = {{\bf{x}}^t}}^T.
%\end{split}
%\end{equation}

 \begin{algorithm}
\caption{BLS for a Feasible $\delta $ in the $t$-th iteration}
  \begin{algorithmic}[1]
\State \textbf{Input:} ${{\bf{x}}^{t - 1}}$, $\delta  > 0$, $0 < \rho  < 1$.

\State \textbf{Repeat:}

\State \quad ${{\bf{x}}^{t}} = {\cal B}\left\{ {{{\bf{x}}^{t - 1}} - \delta {\nabla _{{{\bf{x}}^t}}}f({\bf{x}}),{\cal C}} \right\}$.

\State \quad If $f({{\bf{x}}^{t}}) > f({{\bf{x}}^{t - 1}}) - \delta \left\| {{\nabla _{{{\bf{x}}^{t - 1}}}}f({\bf{x}})} \right\|_2^2$, update $\delta  \leftarrow \rho \delta $.

\State \textbf{End}

\State \textbf{Until:} $f({{\bf{x}}^{t}}) \le f({{\bf{x}}^{t - 1}}) - \delta \left\| {{\nabla _{{{\bf{x}}^{t - 1}}}}f({\bf{x}})} \right\|_2^2$.
  \end{algorithmic}
\end{algorithm}

 \begin{algorithm}
\caption{The Overall Algorithm for Solving (P2)}
  \begin{algorithmic}[1]
\State \textbf{Input:} $t = 1$, ${{\bf{x}}^1} \in {\cal C}$.

\State \textbf{Repeat:}

\State \quad Perform eigenvalue decomposition on ${\left[ {\bf{Z}} \right]_{{\bf{x}} = {{\bf{x}}^t}}}$ and

\noindent \quad compute ${\left[ {\left\{ {\partial {\bf{Z}}/\partial {x_n}} \right\}_{n = 1}^N} \right]_{{\bf{x}} = {{\bf{x}}^t}}}$ to obtain ${\nabla _{{{\bf{x}}^t}}}f({\bf{x}})$.

\State \quad Determine a feasible $\delta $ based on Algorithm 1;

\State \quad $t \leftarrow t + 1$;

\State \quad Update ${{\bf{x}}^{t}} = {\cal B}\left\{ {{{\bf{x}}^{t - 1}} - \delta {\nabla _{{{\bf{x}}^{t - 1}}}}f({\bf{x}}),{\cal C}} \right\}$.

\State \textbf{End}

\State \textbf{Until:} $\left| {f({{\bf{x}}^{t}}) - f({{\bf{x}}^{t - 1}})} \right| \le \tau $.
  \end{algorithmic}
\end{algorithm}

%\subsection{Determining the feasible step size}
\textit{\underline{B. Determining the feasible step size:}} In the PGD method, a correct setting for the step size in each iteration is important for realizing convergence. Specifically, the feasible $\delta $ in each iteration should satisfy $\delta  \le 1/{L_{\bf{x}}}$, where ${L_{\bf{x}}}$ is a Lipschitz constant for ${\nabla _{\bf{x}}}f({\bf{x}})$, which satisfies ${\left\| {{\nabla _{\bf{x}}}f({\bf{x}}) - {\nabla _{{\bf{x'}}}}f({\bf{x}})} \right\|_2} \le {L_{\bf{x}}}{\left\| {{\bf{x}} - {\bf{x'}}} \right\|_2}$, $\forall {\bf{x}},{\bf{x'}} \in {\cal C}$ [15]. Since the structure of ${{\nabla _{\bf{x}}}f({\bf{x}})}$ is much complex, generally ${L_{\bf{x}}}$ is difficult to determine. Based on this fact, we can instead exploit the backtracking line search (BLS) [16] to find a feasible $\delta $. The details are shown in Algorithm 1, where $\rho $ denotes the shrinking factor.

%\subsection{Determining the projection function ${\cal B}\left\{  \cdot  \right\}$}
\textit{\underline{C. Determining the projection function ${\cal B}\left\{  \cdot  \right\}$:}} Recall that the projection function mainly ensures that $N$ FAs only move in their respective feasible regions. Therefore, according to the rule of nearest distance, ${\cal B}\left\{ {{{\bf{x}}^{t + 1}},{\cal C}} \right\}$ can be determined as
%\begin{equation}
%\setcounter{equation}{19}
%\begin{split}{}
%&{\cal B}\left\{ {{{\bf{x}}^{t + 1}},C} \right\}\\
% \triangleright  & \left[ {x_i^{t + 1} = \left\{ {\begin{array}{*{20}{c}}
%{{F_i}}&{x_i^{t + 1} < {F_i}}\\
%{x_i^{t + 1}}&{{F_i} < x_i^{t + 1} < {G_i}}\\
%{{G_i}}&{x_i^{t + 1} > {G_i}}
%\end{array}} \right.} \right]_{i = 1}^N.
%\end{split}
%\end{equation}
\begin{equation}
\setcounter{equation}{19}
\begin{split}{}
&{\cal B}\left\{ {{{\bf{x}}^{t + 1}},C} \right\} \triangleright x_i^{t + 1} = \min \left( {\max ({F_i},x_i^{t + 1}),{G_i}} \right).
\end{split}
\end{equation}

%\subsection{The overall algorithm, complexity analysis and comparison}
\textit{\underline{D. The algorithm, complexity analysis and comparison:}} The overall setups for solving problem (P2) are summarized in Algorithm 2, where $\tau $ denotes the prescribed accuracy. Generally, the PGD based minimization may lead to sightly different total transmit power for different initialization ${{\bf{x}}^1}$ and step-sizes $\delta $. This is mainly because the PGD may converge to a local minimum of the objective, which is an unavoidable phenomenon arising in non-convex optimization problems. Nevertheless, this phenomenon can be well solved by randomly generating numerous different ${{\bf{x}}^1}$ and then selecting the one which produces the minimum power.

\textit{Complexity Analysis:} To simplify the analysis while still capturing the complexity of Algorithm 2, we here focus on the number of complex multiplications required in each iteration. Specifically, the complexity of the eigenvalue decomposition for ${\left[ {\bf{Z}} \right]_{{\bf{x}} = {{\bf{x}}^t}}}$ is about ${\cal O}({M^3})$ [9]. Further, calculating $\sum\nolimits_{i = 1}^M { - {{\left[ {{{\bf{V}}^{ - 1}}} \right]}_{i,:}}\frac{{\partial {\bf{Z}}}}{{\partial {x_n}}}{{\left[ {\bf{V}} \right]}_{:,i}}/\lambda _i^2\left\{ {\bf{Z}} \right\}} $, $\forall n = 1,...,N$, requires ${\cal O}({M^2})$ complex multiplications, leading to the complexity of computing ${\nabla _{{{\bf{x}}^t}}}f({\bf{x}})$ as ${\cal O}({M^2}N)$. In addition, the complexity of finding a feasible $\delta $ is about ${\cal O}({T_{{\rm{inner}}}}N)$, where $N$ is the complexity of computing $\delta {\nabla _{{{\bf{x}}^t}}}f({\bf{x}})$ in step 3 of Algorithm 1, and ${T_{{\rm{inner}}}}$ is the maximum number of iterations for BLS. Hence, the total complexity of Algorithm 2 is about
\begin{equation} \nonumber
\begin{split}{}
{\cal O}\left( {{T_{{\rm{outer}}}}\left( {{M^3} + {M^2}N + {T_{{\rm{inner}}}}N} \right)} \right),
\end{split}
\end{equation}
where ${{T_{{\rm{outer}}}}}$ is the maximum number of iterations for repeatedly implementing steps 3-5 in Algorithm 2.

\textit{Complexity Comparison:} As a comparison, if the original definition based method [9] is exploited to compute the gradient, i.e.,
\begin{equation}
\begin{split}{}
&\frac{{\partial f({\bf{x}})}}{{\partial {x_n}}}{|_{{\bf{x}} = {{\bf{x}}^t}}} = \mathop {\lim }\limits_{\varepsilon  \to 0} \frac{{f(x_1^t,...,x_n^t + \varepsilon ,...,x_N^t) - f({{\bf{x}}^t})}}{\varepsilon },
\end{split}
\end{equation}
 the corresponding complexity will become larger. Specifically, given ${{{\bf{x}}^t}}$ and $\varepsilon $, using the eigenvalue decomposition to obtain ${f(x_1^t,...,x_n^t + \varepsilon ,...,x_N^t)}$ for all $n = 1,...,N$ requires a complexity of ${\cal O}(N{M^3})$. Similarly, the complexity of obtaining ${f({{\bf{x}}^t})}$ is ${\cal O}({M^3})$. Therefore, the complexity of obtaining ${\nabla _{{{\bf{x}}^t}}}f({\bf{x}})$ is about ${\cal O}((N + 1){M^3})$, and then the total complexity of Algorithm 2 becomes
 \begin{equation} \nonumber
{\cal O}\left( {{T_{{\rm{outer}}}}\left( {{M^3} + {M^3}N + {T_{{\rm{inner}}}}N} \right)} \right),
\end{equation}
which is clearly higher than the complexity of Algorithm 2 in this work, especially when $M$ is large. We compare the above two complexities versus $M$ in Fig. 3 for better illustration, where we set ${T_{{\rm{outer}}}} = {T_{{\rm{inner}}}} = 10$ and $N = 30$.
%\subsection{Performance bound}
%By exploiting the fact that the function $1/x$ is convex w.r.t. $x$, we can derive that
%\begin{equation}
%\begin{split}{}
%&f({\bf{x}}) = \sum\nolimits_{i = 1}^M {\frac{1}{{{\lambda _i}({\bf{Z}})}}}  \ge \frac{M}{{\sum\nolimits_{i = 1}^M {{\lambda _i}({\bf{Z}})} }}\\
% =& \frac{M}{{{\rm{tr}}\left( {\bf{Z}} \right)}} \mathop  = \limits^{(d)} \frac{{M{\sigma ^2}}}{{N\sum\nolimits_{i = 1}^M {\frac{1}{{{\varepsilon _i}}}} }} \buildrel \Delta \over = {f_{{\rm{LB}}}},
%\end{split}
%\end{equation}
%where $\mathop  = \limits^{(d)} $ is established since by focusing on (16), it has ${\left[ {\bf{Z}} \right]_{i,i}} = \frac{N}{{{\varepsilon _i}{\sigma ^2}}}$ and then ${\rm{tr}}\left( {\bf{Z}} \right) = \sum\nolimits_{i = 1}^M {{{\left[ {\bf{Z}} \right]}_{i,i}}}  = N\sum\nolimits_{i = 1}^M {\frac{1}{{{\varepsilon _i}{\sigma ^2}}}} $.

%Equation (20) provides a globally lower bound for the total transmit power of all $M$ users, which is inversely proportional to number of antennas at the BS, but is proportional to the noise power and the minimum rate requirement at each user. From subsequent simulations we will observe that via careful optimization for the positions of the MA array ${\bf{x}}$, the total transmit power can be reduced  significantly and approach ${f_{{\rm{LB}}}}$ as far as possible.

 \begin{figure}
 %\vspace{-1pt}
\centering
\includegraphics[width=6cm]{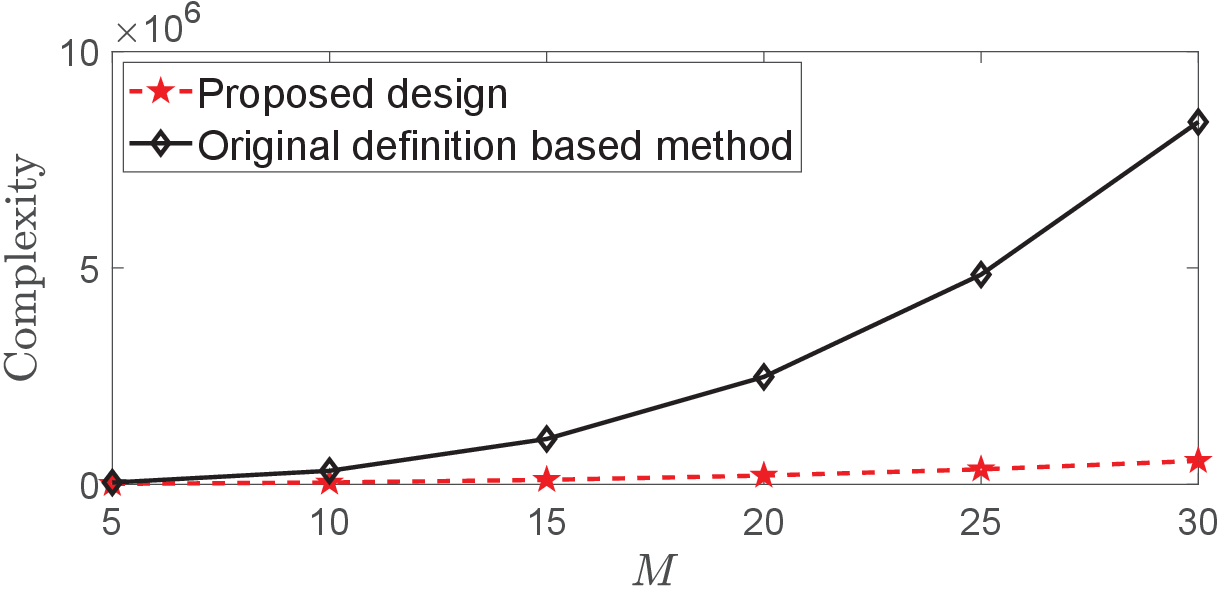}
\captionsetup{font=small}
\caption{Computational complexity of the proposed design and the original definition based method.} \label{fig:Fig1}
\end{figure}

  \begin{figure}
 %\vspace{-1pt}
\centering
\includegraphics[width=6cm]{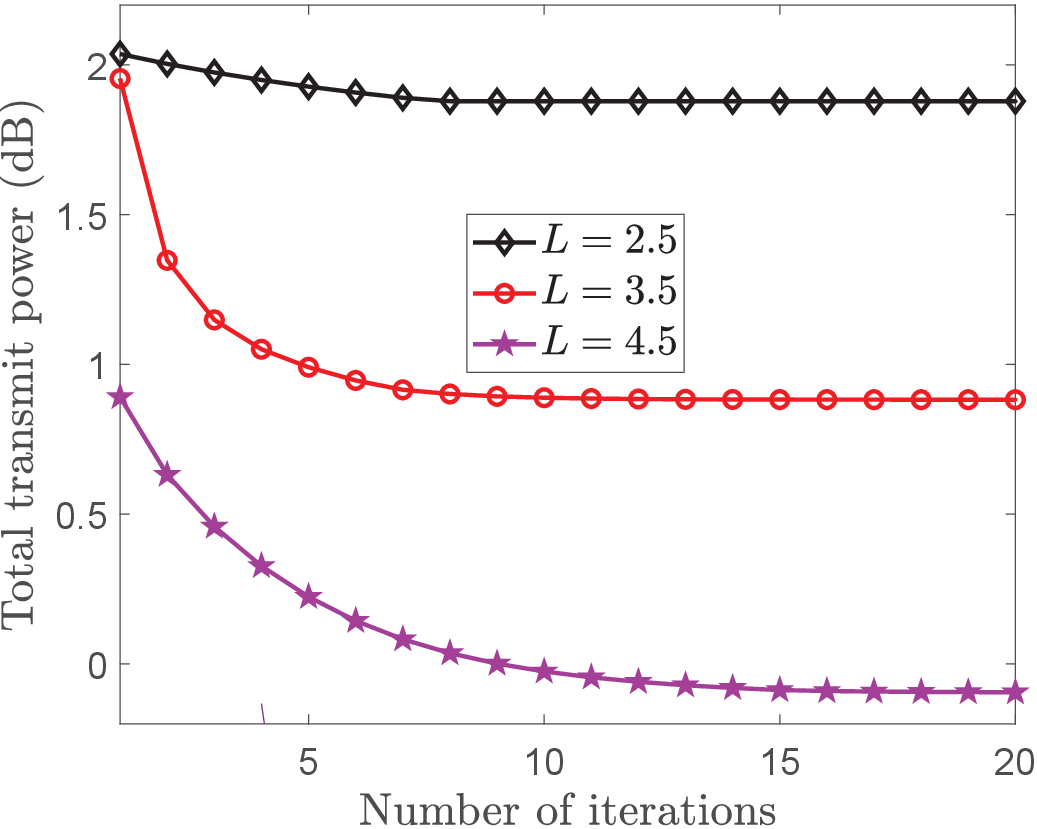}
\captionsetup{font=small}
\caption{The convergence behavior of the proposed PGD method.} \label{fig:Fig1}
\end{figure}

\begin{figure*}
%\vspace{-18pt}
\centering

\begin{minipage}{5.4cm}
\includegraphics[width=5.4cm]{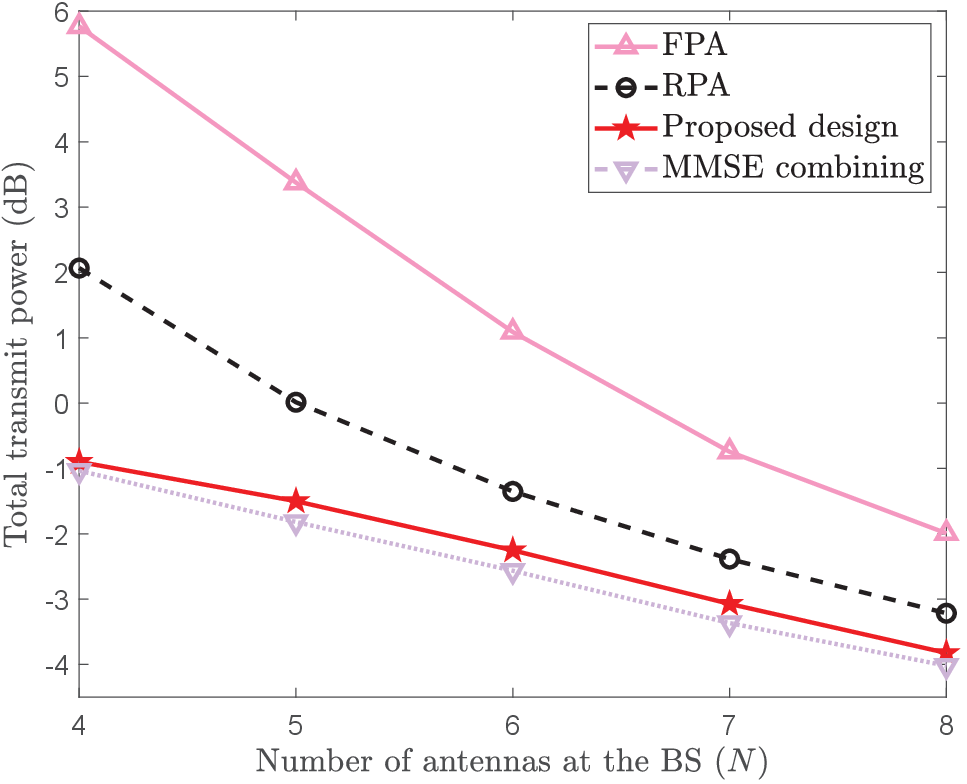}
\centering
%\vspace{-5pt}
\subfigure{(a)}

\end{minipage}
\begin{minipage}{5.4cm}
\includegraphics[width=5.4cm]{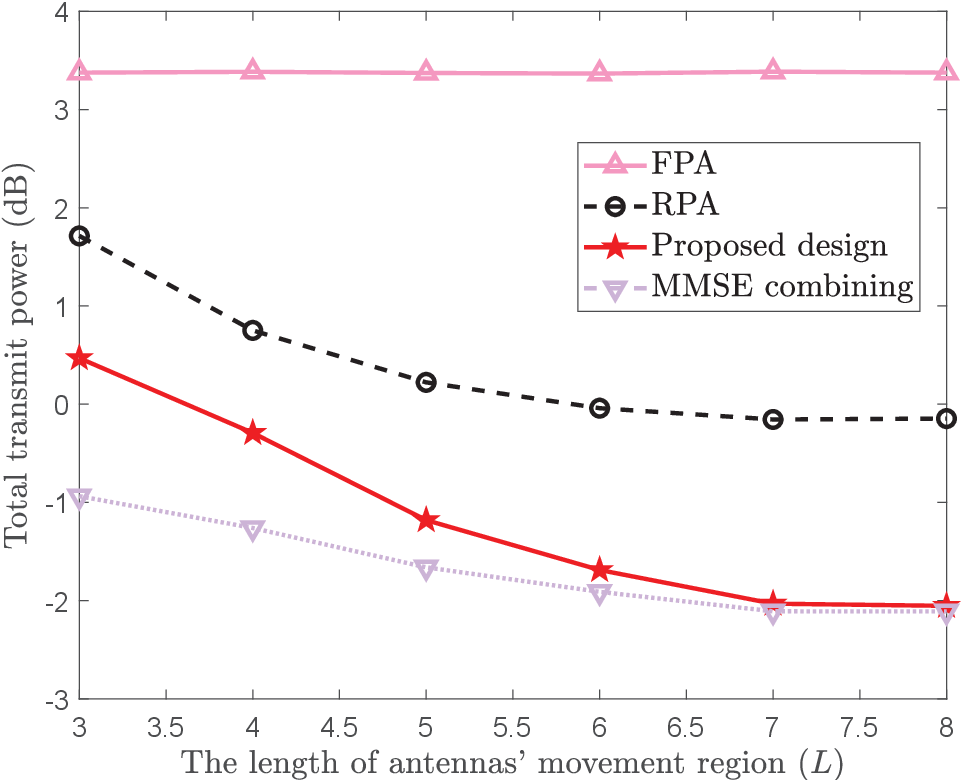}
\centering
%\vspace{-5pt}
\subfigure{(b)}

\end{minipage}
\begin{minipage}{5.4cm}
\includegraphics[width=5.4cm]{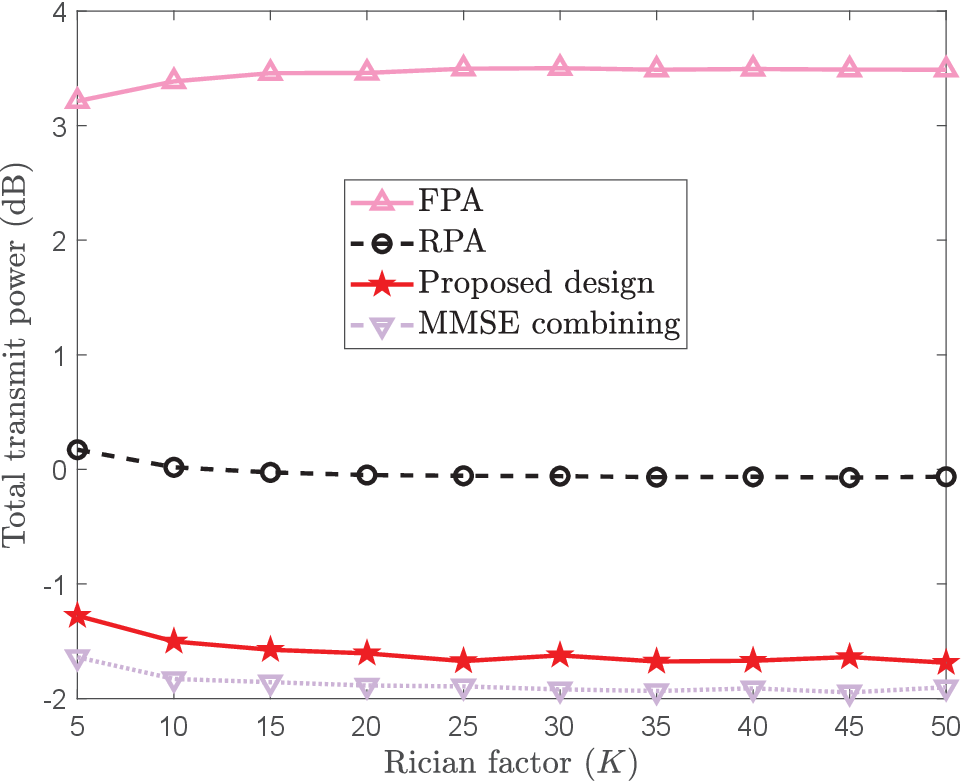}
\centering
%\vspace{-5pt}
\subfigure{(c)}

\end{minipage}
%\vspace{-3pt}
\caption{Average total transmit power versus (a) $N$; (b) $L$; (c) $K$.}
%\vspace{-7pt}
\end{figure*}

% \begin{figure}
% %\vspace{-1pt}
%\centering
%\includegraphics[width=6cm]{ver_N.eps}
%\captionsetup{font=small}
%\caption{Total transmit power versus number of antennas at the BS.} \label{fig:Fig1}
%\end{figure}

 \section{Simulation Results}
In this section, we present numerical results to demonstrate the effectiveness of the proposed design over the general Rician fading, in which the channel vector between the BS and ${\rm{U}}_i$ is ${\widehat {\bf{h}}_i}({\bf{x}}) = \sqrt {K/(K + 1)} {{\bf{h}}_i}({\bf{x}}) + \sqrt {1/(K + 1)} {\widetilde {\bf{h}}_i}$, where $K$ is the Rician factor, ${{\bf{h}}_i}({\bf{x}})$ is given in (1), and each element of ${\widetilde {\bf{h}}_i} \in {{\mathbb{C}}^{N \times 1}}$ is i.i.d. complex Gaussian distributed with zero mean and unit variance. Under this setup, optimal ${\bf{x}}$ (denoted as ${{\bf{x}}^{{\rm{LoS}}}}$) is still obtained based on statistical AoAs, while the objective of total transmit power becomes ${\mathbb{E}}\left[ {\sum\nolimits_{i = 1}^M {\frac{1}{{{\lambda _i}\left\{ {{{\bf{\Omega }}^{ - 1}}\widehat {\bf{H}}{{({{\bf{x}}^{{\rm{LoS}}}})}^H}\widehat {\bf{H}}({{\bf{x}}^{{\rm{LoS}}}})} \right\}}}} } \right]$, with $\widehat {\bf{H}}({{\bf{x}}^{{\rm{LoS}}}}) = \left[ {{{\widehat {\bf{h}}}_1}({{\bf{x}}^{{\rm{LoS}}}}),...,{{\widehat {\bf{h}}}_M}({{\bf{x}}^{{\rm{LoS}}}})} \right]$. For convincing comparisons, we further consider three widely used benchmarks:
\begin{itemize}
		\item RPA: The line segment of length $L$ is quantized into $2L + 1$ discrete locations with equal-distance $0.5\lambda $, and $N$ out of these $2L + 1$ locations are optimally selected for antenna positions.
		\item FPA: Each antenna has a fixed position, i.e., ${x_i} = (i - 1){d_{\min }}$.
\item Minimum mean square error (MMSE) combining: The BS will exploit MMSE combining to detect multiple signals, where positions of all antennas are optimized employing the method in [9], but base on statistical AoAs.
	\end{itemize}

For the system parameters, we set the minimum distance between any two adjacent FAs as ${d_{\min }} = 0.5\lambda $, and without prejudice to the conclusion, $\lambda $ is set to 1 for simplification. We consider $M = 3$ users and the AoAs are ${\theta _1} = \pi /16$, ${\theta _2} = \pi /10$ and ${\theta _3} = \pi /2$, respectively. In addition, the noise power is set as ${\sigma ^2} = 1$ for normalizing the large-scale channel fading power.

Fig. 4 first illustrates the convergence behavior of our proposed design under the LoS channels and for the case of $N = 4$ and ${r_i} = 1$, $\forall i = 1,2,3$. Corresponding to different $L = 2.5, 3.5, 4.5$, the initial condition for the iteration is set as ${{\bf{x}}^1} = {[0,L/3,2L/3,L]^T}$. As we can observe, the total transmit power of all users rapidly converges to a constant within dozens of iterations. Therefore, the proposed design is computationally efficient which may be suitable for the practical implementation.

% \begin{figure}
%% \vspace{-1pt}
%\centering
%\includegraphics[width=6cm]{ver_r.eps}
%\captionsetup{font=small}
%\caption{Total transmit power versus the minimum rate requirement.} \label{fig:Fig1}
%\end{figure}
%
% \begin{figure}
% %\vspace{-1pt}
%\centering
%\includegraphics[width=6cm]{ver_L.eps}
%\captionsetup{font=small}
%\caption{Total transmit power versus the length of the line segment for the MAs' movement.} \label{fig:Fig1}
%\end{figure}

Fig. 5(a) compares the total transmit power of four schemes with respect to (w.r.t.) number of transmit antennas at the BS ($N$) for the case of ${r_i} = 1$, $\forall i = 1,2,3$ and $K = 10$. We can observe that: i) as $N$ increases, the BS can better distinguish signals in different directions and achieve higher reception gains, which in turn allows the users to transmit their signals with less power; ii) compared to FPA and RPA, the proposed design can optimally exploit the additional spatial DoF, so that the resulting total transmit power can be minimized; iii) as $N$ increases, the performance gap between RPA and the proposed design decreases. The reason is that when $L$ is fixed, each antenna can just move in a smaller region when $N$ increases, which implies that there may be not much performance difference from discrete positions selection in RPA or optimal FAs' movements in the proposed design; iv) as reported in [9], due to the more powerful detection ability, MMSE combining outperforms the proposed design slightly.

Fig. 5(b) shows the total transmit power w.r.t. the span of FAs' movement ($L$) for the case of ${r_i} = 1$, $\forall i = 1,2,3$ and $K = 10$, from which it is observed that when $L$ increases, the total transmit power of RPA, the proposed design and MMSE combining first becomes smaller and then converges to a constant. This phenomenon reveals that it is not necessary to expand $L$ indefinitely and only a limited span is enough to achieve the optimal performance.

Finally, Fig. 5(c) shows the total transmit power w.r.t. the Rician factor $K$ for the case of ${r_i} = 1$, $\forall i = 1,2,3$, $N = 5$ and $L = 5.5$, from which it is observed that no matter whether $K$ is large (the LoS condition is dominate for each channel between the user and the BS) or small (the random Rayleigh fading is dominate for each channel between the user and the BS), our proposed design with statistical AoAs always achieves pretty good performance compared to FPA and RPA, indicating that our proposed design is not sensitive w.r.t. random fading components in Rician channels.

\vspace{-10pt}
\section{Conclusion}
This letter considers multiuser uplink communication supported by the FAs-enabled base station, which exploits zero-forcing receivers to decode multiple signals. The objective is to optimize the FAs' positions at the BS, to minimize the total transmit power of all users subject to the minimum rate requirement. We develop a projected gradient descent method to iteratively find a locally optimal solution, at significantly reduced complexity compared to state of the art since a closed-form gradient is derived successfully. Results show the performance superiority of our proposed design compared to several benchmarks.

\end{document}